\documentclass[12pt,preprint]{aastex}

\shorttitle{NIR Polarimetry of Orion Nebula}
\shortauthors{Tamura et al.}

\begin{document}

\title{NEAR INFRARED POLARIZATION IMAGES OF THE ORION NEBULA}

\author{M. Tamura\altaffilmark{1, 2}, R. Kandori\altaffilmark{1}, 
N. Kusakabe\altaffilmark{2}, Y. Nakajima\altaffilmark{1}, J. Hashimoto\altaffilmark{1, 3}}
\author{C. Nagashima\altaffilmark{4}, T. Nagata\altaffilmark{5}, T. Nagayama\altaffilmark{5}, 
H. Kimura\altaffilmark{6}, T. Yamamoto\altaffilmark{6},}
\author{J. H. Hough\altaffilmark{7}, P. Lucas\altaffilmark{7}, 
A. Chrysostomou\altaffilmark{7}, J. Bailey\altaffilmark{8}}
\author{{\it Astrophysical Journal Letters, in press}}


\altaffiltext{1}{National Astronomical Observatory of Japan, Osawa, Mitaka, Tokyo 
181-8588, Japan; hide@subaru.naoj.org}
\altaffiltext{2}{Department of Astronomical Science, Graduate University for 
Advanced Studies (Sokendai), Osawa, Mitaka, Tokyo 181-8588, Japan}
\altaffiltext{3}{Department of Physics, Tokyo University of Science, 1-3, Kagurazaka, 
Sinjuku-ku, Tokyo 162-8601, Japan}
\altaffiltext{4}{Department of Astrophysics, Faculty of Sciences, Nagoya University, 
Chikusa-ku, Nagoya 464-8602, Japan}
\altaffiltext{5}{Department of Astronomy, Kyoto University, Sakyo-ku, Kyoto 606-8502, Japan}
\altaffiltext{6}{Institute of Low Temperature Science,
Hokkaido University, Sapporo 060-0819, Japan}
\altaffiltext{7}{Centre for Astrophysics Research, University of Hertfordshire, 
Hatfield, Herts AL10 9AB, UK}
\altaffiltext{8}{Australian Centre for Astrobiology, Macquarie University, NSW, Australia}

\begin{abstract}
Wide-field ($\sim$8$'$ $\times$ 8$'$) and deep near-infrared ($JHKs$ bands) polarization 
images of the Orion Nebula are presented. 
These data revealed various circumstellar structures as infrared 
reflection nebulae (IRN) around young stellar objects (YSOs), both massive and low-mass. 
We found the IRN around both IRc2 and BN to be very extensive,
suggesting that there might be two extended ($>$0.7 pc) bipolar/monopolar IRN
in these sources.
We discovered at least 13 smaller-scale ($\sim$0.01-0.1 pc) IRN around less-massive YSOs 
including the famous source $\theta${$^{2}$} Ori C.
We also suggest the presence of many unresolved ($<$690 AU) systems around low-mass YSOs 
and young brown dwarfs showing possible intrinsic polarizations. 
Wide-field infrared polarimetry is thus demonstrated to be a powerful technique 
in revealing
IRN and hence
potential disk/outflow systems
among high-mass to substellar YSOs.
\end{abstract}


\section{INTRODUCTION}

The Orion Nebula and its associated population of stars are amongst the best studied objects 
in the sky. 
As a component of the ridge of molecular cloud (OMC-1) extending north-south, 
M42 is the HII region excited by the Trapezium, a group of massive (OB-type) young stars, 
near the surface of the molecular cloud. 
Of the Trapezium stars, $\theta${$^1$} Ori C is the most dominant photo-ionizing source. 
One arcmin to the north-west on the sky are the two massive YSOs, IRc2 and BN, 
whose masses are suggested to be 25 and {$\gtrsim$}7 {$M$}$_{\sun}$, respectively 
(see Genzel \& Stutzki 1989; Jiang et al. 2005). 
In fact IRc2 is
a cluster of sources seen at infrared wavelengths, with additional nearby
sources detected at different wavelengths (see e.g., Beuther et al. 2004)
such as the radio source (source I) and the submillimetre source (SMA1).  
For convenience we refer to the
sources responsible for the major outflows as the ``IRc2 prtostars''.
At least two outflows are known to originate from the region near the IRc2 protostars; 
one high-velocity outflow in the SE-NW direction, observed in radio molecular lines, 
and in optical and near-infrared shocked lines 
(Allen \& Burton 1993; Chernin \& Wright 1996), 
and one lower-velocity outflow in the NE-SW direction best seen in thermal SiO and H$_{2}$O 
maser emission, as well as some H$_{2}$ bow shocks 
(Genzel \& Stutzki 1989; Chrysostomou et al. 1997).
The driving sources of these outflows are uncertain.
Situated to the south is the Bright Bar, which corresponds to the tangential region 
of the wall of the ionization front sculpting OMC-1. 
Distributed over the entire nebula is the Orion Nebular Cluster (ONC; aka. Trapezium cluster), 
which is composed of some 3500 low-mass YSOs and about 150 stars with protoplanetary disks 
revealed by $HST$ imaging (O'Dell 2001). 

Many of these features are clearly seen with near-infrared intensity imaging 
(see e.g., Hillenbrand \& Carpenter 2000; see also Figure~1). 
Near-infrared wavelengths are useful because of a combination of the low temperature 
of the YSOs and the extensive dust-extinction over the star forming region. 
However, near-infrared radiation from massive star forming regions is varied in its origin: 
not only direct radiation from YSOs but also gas free-free continuum emission, 
dust thermal emission, and dust scattering from various illuminating sources. 
The last scattered component from both diffuse nebulae over the whole region and 
the local environs around YSOs are best studied with polarized radiation.

However, polarimetric studies covering the central few arcmin of M42, where a variety of
activities are seen have been limited so far. Imaging polarimetry
has been conducted either only at optical wavelengths (Pallister et al. 1977)
or toward a small region near IRc2 or BN at near-infrared wavelengths 
(Minchin et al. 1991; Jiang et al. 2005; Simpson et al. 2006).
In this {Letter}, we present for the first time
the near-infrared polarization images of a $\sim$1 pc region of M42
and describe newly detected features seen as IR nebulae on various scales.

\section{OBSERVATIONS AND DATA REDUCTION}
The 1.25 ($J$ band), 1.63 ($H$ band), and 2.14 ($Ks$ band) $\mu$m imaging polarimetry 
of M42 and a sky-field were obtained simultaneously with the SIRIUS camera
(Nagayama et al. 2003)
and its polarimeter on the 1.4-m IRSF telescope in South Africa, 
on the night of 2005 December 26. 
The instrument is among the first ones that provide deep and wide-field infrared polarimetric 
images, which can in principle measure polarizations of all the 2MASS-detected sources 
within a field-of-view of 7$\farcm$7 $\times$ 7$\farcm$7 in the $JHKs$ bands simultaneously with 1\% 
polarization accuracy.
See Kandori et al. (2006) for the details of the polarimeter. 

The total integration time per wave plate position was 900 s and 
the resultant stellar seeing size was 1$\farcs$5. 
The pixel scale was 0$\farcs$45. 
The exposures were performed at 
four position angles (PAs) of the halfwave plate, in the sequence  
of PA~= ~0$\arcdeg$, 45$\arcdeg$, 22$\arcdeg$.5, and 
67$\arcdeg$.5 to measure the Stokes parameters. 
After image calibrations in the standard manner using IRAF (dark subtraction, 
flat-fielding with twilight-flats, bad-pixel substitution, and sky subtraction), 
the Stokes parameters $(I,\, Q,\, U)$, the degree of polarization $p$, and 
the polarization angle $\theta$ were calculated (Kandori et al. 2006).

Software aperture polarimetry was carried out for a number of point sources 
in the field of view. The aperture radius was 3 pixels. 
First the point-like sources were selected after the subtraction 
of the smooth nebulous components. 
Then the local background was subtracted using the mean of a circular annulus 
around the source on the original images. 
We rejected all the sources whose photometric errors were greater than 0.1 mag. 
The polarization percentages were debiased (Wardle \& Kronberg 1974).
In this {Letter}, we discuss only $H$ band data for the aperture polarimetry, 
as the extended nebula contamination is less than in the $J$ band and the scattering efficiency 
is higher than in the $Ks$ band.
In the $H$ band image, 498 sources were measured in total.
The full list of the aperture polarimetry at $JHKs$ will be given in a separate paper.

\section{RESULTS AND DISCUSSION}
\subsection{Polarized Intensity Images of M42 and Large Scale Infrared Nebulae}
The outflows produced by both massive and low-mass YSOs are or were powerful 
enough to open a cavity within their parent cloud or core, 
in a direction that tends to be perpendicular to dense disk-like structures around the star. 
The radiation from the star preferentially escapes in the polar direction of the cavity, 
and is then scattered by the dust at the wall of the cavity, forming so-called polarized 
infrared reflection nebulae (IRN; see e.g., Hodapp 1984; Sato et al. 1985; 
Weintraub, Goodman, \& Akeson 2000)

Figure~1 shows the three-color composite, intensity and polarized intensity images 
in the $J$, $H$, and $Ks$ bands. 
The IRN revealed by our polarization images is a signature of circumstellar structures of 
various scales (from $\sim$1 pc down to less than the seeing size, $<$690 AU). 
The infrared polarization has been used to derive the morphology of the IRN associated 
with past or present outflow phenomena 
(Tamura et al. 1991; Sato et al. 1985; Hodapp 1984), 
either via spatially resolving the IRN or via detecting significant levels of integrated 
(i.e., unresolved) polarization. 

 
Our polarization images reveal a new picture of the geometry in this region. 
The most dominant feature is the bipolar IRN centered on the IRc2 protostars, 
seen as red nebulae extending predominantly to the east and west in the color composite image 
(note that an arc to the north belongs to another IRN illuminated by the BN object 
as described below) and the $Ks$ band polarization vector map in Figure~2
(see also a sketch in Figure~3).
This most likely traces the cavity carved out by the molecular outflow from the IRc2 protostars. 
First, the highly polarized (up to $p$ $\sim$ 40 \% at $Ks$) IRN is more extended 
than previously thought; 
the extent of the outflow traced by our polarization images is $\sim$0.7 pc east-west, 
while the extent of the H$_{2}$ outflow has been suggested to be $\sim$0.3 pc to the NNW 
and only $\sim$0.1 pc to the west (Allen \& Burton 1993; Schultz et al. 1999). 
Therefore, the molecular cloud near the IRc2 protostars might already have had 
a cavity formed with a size-comparable to the diameter of the ionized region around the Trapezium. 
Second, the morphology is different between the H$_{2}$ emission and the IRN; 
in H$_{2}$ emission the well known ``fingers'' (Allen \& Burton 1993) most prominently extend 
to the NW, while in polarized emission the extension is to the west. 
The western nebula clearly shows a V-shaped cavity structure with a wide opening angle 
($\sim$50$\arcdeg$), which is typical for IRN associated with molecular outflows 
(Tamura et al. 1991; Staude \& Els{\"a}sser 1993). 
This, together with the extension to the east, makes the total outflow quite symmetric 
around the IRc2 protostars in both direction and size. 
Note that the cavity direction is perpendicular to the dense ridge of OMC-1. 
Once UV radiation from the IRc2 protostars becomes powerful enough, 
the cavity evacuated by the outflow might become a circular-symmetric HII region, 
as is seen around the Trapezium.

In addition to the IRN illuminated by the IRc2 protostars, there is another large 
IRN illuminated by BN (Simpson et al. 2006) as also seen in the polarization vector 
pattern near BN (Figure~2). 
This IRN extends to the north (seen as a reddish arc with its apex 
coincident with BN), 
but no clear counterpart lobe is seen in our images. 
Although the IRN could simply be due to dust in the cloud that is illuminated by BN,
we suggest that the IRN associated with BN might represent an independent outflow, 
whose direction is distinct from that of the IRc2 protostars.
Such a large (0.3 pc) and curved morphology starting from BN
whose direction is consistent with the polar region of 
the compact polarization disk (Jiang et al. 2005) 
could be carved out by a massive outflow associated with BN.
The misalignment of the BN and IRc2 ``outflows'' is not surprising 
if BN is a ``run-away'' star from the Trapezium region (Tan 2004). 
 
\subsection{Medium Scale Infrared Nebulae around YSOs}
A number of smaller-scale ($\sim$0.01-0.1 pc) IRN are also seen in our polarization images; 
there are at least 13 ``resolved'' IRN (see Figure~3). 
They are around low- to intermediate-mass YSOs. 
Most prominent are those near OMC-1 S and $\theta${$^2$} Ori C. 
The former is seen as a reddish and irregular pattern in the polarized images
(see Hashimoto et al. 2006 for $JKs$ data and more discussion). 
The latter shows a bipolar-pattern in the polarized intensity and 
a centro-symmetric polarization vector pattern.
Although $\theta${$^2$} Ori C is a well-known object and has been observed 
at many times, no circumstellar structure has been suggested previously. 
Note, however, that care must be taken in determining the exact geometry 
of IRN from the polarized intensity images 
in Figure~1 alone, because a combination of circumstellar scattering 
of light from the central star and illumination by the Trapezium stars, 
could systematically change the apparent polarization pattern. 
These (as well as unresolved, highly polarized sources described below)
might serve as the best targets for future high-resolution imaging 
with adaptive optics (e.g., Perrin et al. 2004).

\subsection{Unresolved Infrared Nebulae around Low-Mass YSOs and Young Brown Dwarfs}
There are numerous ``unresolved'' (point-like, $<$690 AU) sources 
detected in our near-infrared images, most of which are low-mass YSOs 
and comprise part of the ONC cluster. 
The circumstellar structure is not resolved in our 1$\farcs$5 resolution images. 
However, software aperture polarimetry of these sources can provide 
geometrical evidence for circumstellar structures 
(Sato et al. 1985; Tamura \& Sato 1989).
Interestingly, 86 (17\%) sources show an aperture polarization larger 
than 6\%, while 58 have a larger $p(H)/(H-K)$ value than that for BN. 
Such a large near-infrared polarization is likely to arise from scattering 
in the small-scale IRN system around each source, rather than originating 
from interstellar or intra-cloud extinction, except for extremely red sources. 
The details of these polarized sources are described elsewhere (N. Kusakabe et al.
in preparation), 
but in this {Letter}, we just point out that
some young brown dwarf candidates, whose masses have been 
estimated to be less than 0.08 {$M$}${_{\odot}}$ from {\it both} photometry and spectroscopy 
(Slesnick, Hillenbrand, \& Carpenter 2004), appear to have intrinsic polarizations 
because their $p(H)/${$\tau$}$(H)$ values are greater than that 
for the dichroic polarization due to extinction in the molecular clouds 
(Jones, Klebe, \& Dickey 1992). 
In particular, the $H$ band polarizations of HC022 (0.05 $M$$_{\sun}$, $A_{V}$ = 2 mag) 
and HC064 (0.03 $M$$_{\sun}$, $A_{V}$ = 4 mag) are 
1.49 $\pm$ 0.06 \% and 5.82 $\pm$ 0.09 \%, respectively, 
and are most likely due to scattering in unresolved IRN. 
It has been suggested that young brown dwarfs are associated with disks 
based on indirect evidence such as excess IR emission 
(e.g., Muench et al. 2001). 
Besides a few examples of disk imaging by $HST$ 
(Bally, O'Dell, \& McCaughrean 2000) 
and a detection of an outflow by spectro-astrometry 
(Whelan et al. 2005), 
our results serve as direct ``geometrical'' evidence for 
IRN associated with disk-outflow 
systems around young brown dwarfs.

\subsection{Polarized Orion Bar}
The polarized radiation at shorter wavelengths (our $J$ band data 
and previous optical data by Pallister et al. 1977) is dominated 
by scattered light from the Trapezium stars rather than that from the IRc2 protostars, 
as evidenced from the polarization vector patterns (Figure 2). 
Our polarization images show exactly where the scattering of this diffuse nebula 
(seen as white) occurs; 
it originates from the boundary between the ionization front and the background molecular cloud, OMC-1. 
This is evidenced by the systematic displacement of the position of the Orion Bar 
seen in the intensity image (corresponding to the ionization front) and 
that seen in the polarized intensity image (corresponding to the scattering region), 
and the very low polarization near the Trapezium. 
The distance between the ionization front and the dust scattering region is $\sim$0.02 pc. 
The typical polarization level in the Bar region is 5-10\% in the $J$ band. 
The scattering dust is mostly concentrated at the boundary because 
the dust is swept-up by stellar radiation pressure from the Trapezium stars 
and slowed down quickly by gas drag in the molecular cloud (Ferland 2001).

%
%
%
%
%
%

\acknowledgments

We appreciate discussions with M. Kurita, S. Sato, and Z. Jiang 
on various aspects of this study.
We also thank an anonymous referee for helpful comments. 
This work is supported by Grants-in-Aid from the Ministry of 
Education, Culture, Sports, Science and Technology of Japan 
(16077101, 16077204), by JSPS (16340061), and by PPARC.

\clearpage

\begin{figure}
\plotone{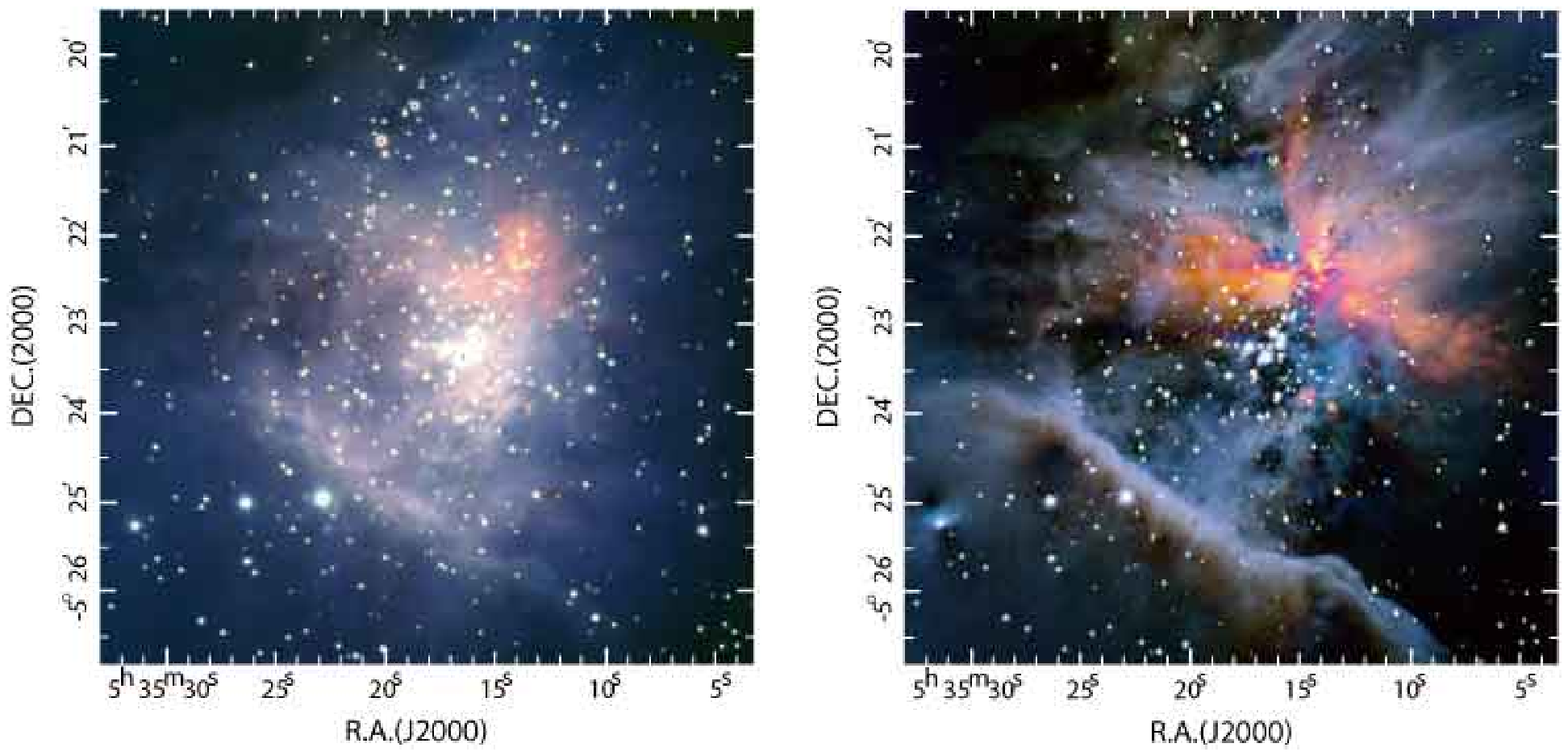}
\caption{
Near-infrared three-color composite images of the Orion Nebula (M42) 
in intensity (Left) and in polarized intensity (Right). 
North is up and east is to the left. 
The images are in logarithmic scale. 
The lowest polarized surface brightness level in the right figure is 
about 19.1-17.6 mag arcsec$^{-2}$ at $J$, $H$, and $Ks$.
\label{fig:m42}} 
\end{figure}

\clearpage

\begin{figure}
\plotone{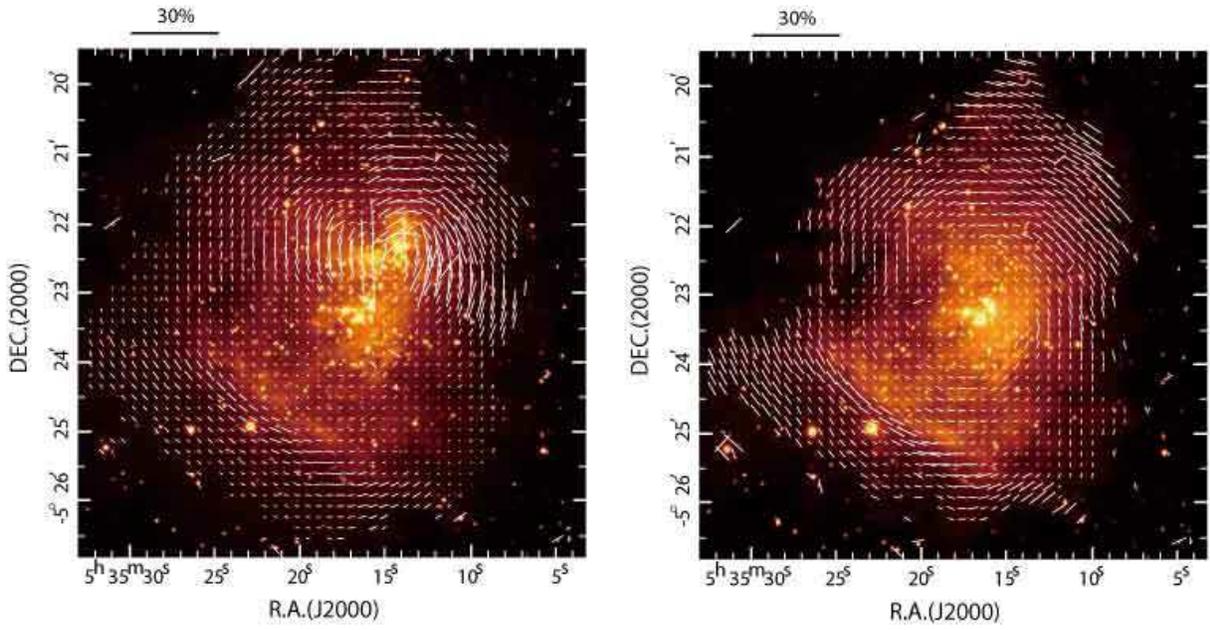}
\caption{
$Ks$ band (Left) and $J$ band (Right) polarization vector maps of M42. 
The vector length is proportional to the degrees of polarization. 
The data are binned at 3 pixels and plotted only every 18 pixels 
for presentation purpose. 
The intensity images are shown in background. 
\label{fig:pol}} 
%
\end{figure}

\clearpage

\begin{figure}
\plotone{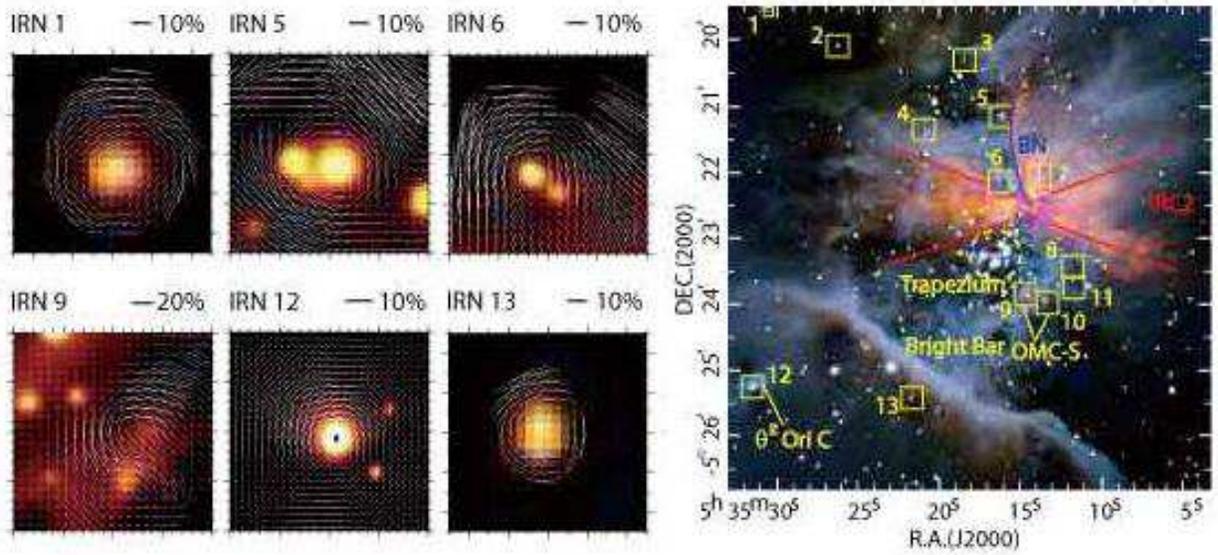} 
\caption{
$H$ band polarization vector maps of six of 
the medium scale IRN (No. 1-13) 
after a subtraction of local background polarization, 
and its identification in the PI image.
The size of each map is 10$''$ $\times$ 10$''$ for IRN1,
20$''$ $\times$ 20$''$ for IRN5, 6, and 9,
40$''$ $\times$ 40$''$ for IRN12, and
10$''$ $\times$ 10$''$ for IRN13.
Sketch of the IRN associated with the IRc2 protostars (red lines) 
and BN (blue line) is also shown. 
Also indicated are 
the Trapezium stars, OMC-1 S, $\theta${$^2$} Ori C, 
and the Bright Bar.
\label{fig:smallIRN}}
\end{figure}

\end{document}